\documentclass{aa}
\def \eg {e.g.}

\def \ie {i.e.}

\def \lcdm {{\hbox{$\Lambda$CDM}}}
\def \omegam {{\hbox{$\Omega_{\rm m}$}}}
\def \omegal {{\hbox{$\Omega_\Lambda$}}}
\def \hzero {{\hbox{$H_0$}}}
\def \arcmin {\hbox{$^\prime$}}
\def \arcsec {\hbox{$^{\prime\prime}$}}
\def \deg {\hbox{$^\circ$}}

\def \msun {\hbox{${\rm M_\odot}$}}

\def \mfive {\hbox{$M_{500}$}}

\newcommand{\kmsmpc }{\mbox{km s$^{-1}$ Mpc$^{-1}$}}

\newcommand{\mujyb }{\mbox{$\mu$Jy beam$^{-1}$}}

\newcommand{\whz }{\mbox{W Hz$^{-1}$}}

\newcommand{\uv }{\textit{uv}}

\newcommand{\wsclean }{\textsc{WSClean}}

\newcommand{\spam }{\textsc{spam}}
\newcommand{\spamE }{Source Peeling and Atmospheric Modeling}

\newcommand{\aoflagger }{\textsc{AOFlagger}}
\newcommand{\dysco }{\textsc{Dysco}}
\newcommand{\dppp }{DP3}
\newcommand{\dpppE }{Default PreProcessing Pipeline}

\newcommand{\xmm }{{\em XMM-Newton}}

\newcommand{\planck }{{\em Planck}}

\newcommand{\xrism }{{\em XRISM}}

\newcommand{\ugmrt }{uGMRT}
\newcommand{\ugmrtE }{upgraded Giant Metrewave Radio Telescope}

\newcommand{\lofar }{LOFAR}

\newcommand{\ska }{SKA}
\newcommand{\skaE }{Square Kilometer Array}

\newcommand{\meerkat }{MeerKAT}

\usepackage{graphicx}
\usepackage{amssymb}
\usepackage{multirow,bigdelim}
\usepackage{txfonts}
\usepackage[export]{adjustbox}
\usepackage{hyperref}
\hypersetup{colorlinks,citecolor=blue,filecolor=black,linkcolor=red,urlcolor=black}
\newcommand{\cl}{G279+39}

\begin{document}

\title{Enhanced radio emission between a galaxy cluster pair}
%\subtitle{}

\authorrunning{Botteon et al.}
\titlerunning{Enhanced radio emission between a galaxy cluster pair}

\author{
Andrea Botteon\inst{\ref{ira}},
Turgay Caglar\inst{\ref{tamu},\ref{leiden}},
Sibel D\"{o}ner\inst{\ref{instanbuluni},\ref{tamu}},
Reinout J. van Weeren\inst{\ref{leiden}},
\and
Krista Lynne Smith\inst{\ref{tamu}}
}

\institute{
INAF -- IRA, via P.~Gobetti 101, 40129 Bologna, Italy \label{ira} \\
\email{andrea.botteon@inaf.it}
\and
George P. and Cynthia Woods Mitchell Institute for Fundamental Physics and Astronomy, Texas A\&M University, College Station, TX, 77843, USA \label{tamu}
\and
Leiden Observatory, Leiden University, PO Box 9513, 2300 RA Leiden, The Netherlands \label{leiden}
\and
Istanbul University, Faculty of Science, Department of Astronomy and Space Sciences, 34116, Istanbul, T\"urkiye \label{instanbuluni}
}

\date{Received XXX; accepted YYY}

\abstract
{Interacting pairs of galaxy clusters offer a unique opportunity to study the properties of the gas residing in the intracluster bridge connecting them. As a consequence of the encounter, both the X-ray and radio emission from the gas are expected to be enhanced by shocks and turbulence, facilitating their detection.}
{PSZ2 G279.79+39.09 is likely an off-axis merging system at $z = 0.29$ with its two main cluster components observed at a projected distance of $\sim$1.3 Mpc. In this paper, we investigate the presence of diffuse radio emission associated with the system.}
{We observed this cluster pair with the \meerkat\ UHF band (544--1088 MHz) for 7.5~h and with the \ugmrt\ band 3 (300--500 MHz) for 8~h. These are the first targeted radio observations of this system.}
{We discover diffuse synchrotron emission in the system, with indication of enhanced emission in the region bridging the cluster pair. The detection is based on the \meerkat\ UHF data, while the  \ugmrt\ band 3 observation does not allow us to derive a stringent limit on the spectral index of the source. This emission is likely generated by the turbulence injected as a consequence of the cluster-cluster encounter. However, the study of its physical properties is limited by the observations currently available on the target. If the two clusters have not yet collided, this emission would resemble the radio bridges observed in A399-A401 and A1758N-S.}
{As other systems with multiple cluster components studied in recent years, the analyzed cluster pair represents an appealing target to investigate the presence of nonthermal phenomena beyond the well-studied denser regions of the intracluster medium. While in this work we presented a new detection, our analysis underlines the need for multi-band observations to fully understand these kinds of sources.}

\keywords{galaxies: clusters: intracluster medium -- galaxies: clusters: general -- galaxies: clusters: individual: PSZ2 G279.79+39.09 -- radiation mechanisms: nonthermal}

\maketitle
%
%-------------------------------------------------------------------

\section{Introduction}

Cluster mergers stand as the most energetic events in the Universe since the Big Bang. Upon collision, they inject turbulence and shocks into the intracluster medium (ICM), often giving rise to expansive synchrotron sources such as radio halos and relics \citep[\eg][]{vanweeren19rev}. These cluster-wide nonthermal phenomena are nowadays observed in numerous massive clusters undergoing merger events \citep[\eg][]{knowles22, botteon22dr2, duchesne24psz2}. Halos and relics allow us to investigate the complex mechanisms at play within the ICM, where kinetic energy of large-scale motions can be channeled into electromagnetic fluctuations and particle acceleration at smaller scale \citep[\eg][]{brunetti14rev}. These radio sources serve also as probes for studying the astrophysical processes associated with galaxy cluster mergers and, in general, contribute to our understanding of the formation and evolution of the large-scale structure in the Universe. \\
\indent
In recent years, significant interest has been devoted to the study of physically bound cluster pairs or systems with multiple cluster components. Observing closely spaced clusters, where individual components can be clearly identified (\eg\ as distinct X-ray-emitting clumps), gives us the opportunity to study the properties of the intracluster gas, beyond the dense cores that are more easily accessible to observations \citep[\eg][]{werner08, kato15, bulbul16, akamatsu16, caglar17, caglar18, kaya19, alvarez22, mirakhor22a2029a2033, omiya23, migkas25double}. Indeed, as a consequence of the encounter, both the X-ray and radio emissions in the bridges of matter connecting the clusters are expected to be boosted \citep{vazza19}. The detection of the radio bridges connecting the cluster pairs A399-A401 \citep{govoni19} and A1758N-S \citep{botteon18a1758, botteon20a1758} has demonstrated that relativistic particles can be accelerated by mechanisms operating on scales larger than the individual clusters, in physical regimes that are poorly explored. One hypothesis is that these radio bridges are powered by novel particle acceleration mechanisms triggered by turbulence generated in the bridges of matter connecting dynamically active clusters \citep{brunetti20}. This model predicts that the radio bridge emission is volume filling and characterized by steep synchrotron spectra (\ie\ $\alpha \gtrsim 1.3$, where $S_\nu\propto\nu^{-\alpha}$). This scenario seems supported by the follow-up observations of A399-A401 and A1758N-S \citep{botteon20a1758, dejong22, nunhokee23, pignataro24lba}. It is however worth noting that reports of diffuse radio emission in clusters with two or more components are not limited to the cases of A399-A401 and A1758N-S. A number of targeted observations using various \skaE\ (\ska) pathfinder and precursor instruments have revealed diffuse synchrotron sources located in-between other clusters \citep{gu19, duchesne21atypical, hoeft21, venturi22, pignataro24a2061, hu25bridge, stuardi25}, while others have resulted in non-detections \citep{botteon19lyra, botteon19a781, bruggen21, pignataro24ugmrt}. The emerging picture is that extended synchrotron sources in these environments are not ubiquitous and can exhibit a broad range of morphologies, sometimes appearing as bridges of radio emission connecting X-ray clumps, and in other cases as filamentary and/or elongated structures. Whether all these sources can be attributed to the same dynamical processes and acceleration mechanisms remains unclear, motivating further theoretical and observational work on systems with multiple cluster components. \\
\indent
PSZ2 G279.79+39.09 (hereafter \cl) was identified in the second \planck\ catalog of Sunyaev-Zel'dovich (SZ) sources \cite{planck16xxii}. It forms a massive ($\mfive = 6.12 \times 10^{14}$ \msun) binary cluster at $z=0.29$, in which its east and west components, identified by their X-ray emission, are separated by a projected distance of $\sim$1.3 Mpc. This system has been overlooked in the literature, lacking a detailed study in any band. We recently analyzed the \xmm\ observation available on this system and characterized the properties of the two subclusters and of the thermal emission between them \citep{doner25sub}. Our analysis suggests that \cl\ is a massive cluster pair exhibiting X-ray asymmetries, indicative of ram-pressure stripping, and high-temperatures in the region connecting the two subclusters, consistent with shock-heated or compressed gas. These features support a scenario in which the two systems are undergoing a merger with a non-zero impact parameter and are observed close to core passage ($\lesssim$0.5~Gyr before or after). However, whether the clusters have already reached or are approaching pericenter remains uncertain, as their projected separation and observed properties are consistent with both phases seen in idealized binary merger simulations \citep{zuhone18catalog}. In this paper, we present new \meerkat\ UHF \citep{jonas09} and \ugmrtE\ \citep[\ugmrt;][]{gupta17ugmrt} band 3 observations of \cl\ with the goal of investigating the presence of diffuse radio emission in the system. In particular, in Section~\ref{sec:reduction}, we describe the data and processing of the radio observations. In Section~\ref{sec:results}, we present the results based on the analysis of our new radio images. In Sections~\ref{sec:discussion} and \ref{sec:conclusions}, we discuss the findings of our work and summarize our conclusions, respectively. \\
\indent
Throughout the paper, we adopted a \lcdm\ cosmology with $\omegal = 0.7$, $\omegam = 0.3$ and $\hzero = 70$ \kmsmpc. This corresponds to a scale conversion factor of 4.350 kpc/arcsec and a luminosity distance of $D_{\rm L} = 1493$ Mpc at the redshift of \cl.

\section{Data reduction}\label{sec:reduction}

\subsection{MeerKAT}

We observed \cl\ with \meerkat\ UHF (544--1088 MHz) for 7.5~h (project code SCI-20241101-AB-02, PI: A. Botteon) on 2025 February 13. The bandwidth of the observation was 544 MHz and was covered with 4096 channels. The observing time includes three 10-min scans on the flux and bandpass calibrators (J0408-6545 and J1939-6342) and a 2-min scan on the gain calibrator J1051-2023 following each passage on the target. \\
\indent
After downloading the data calibrated through the SARAO Science Data Processor (SDP) calibration pipeline, we used the \texttt{facetselfcal.py}\footnote{\url{https://github.com/rvweeren/lofar_facet_selfcal}} script \citep{vanweeren21} to improve the calibration on the field. This script was extensively used to calibrate \lofar\ observations and has been recently adapted and successfully applied also to \meerkat\ data \citep[see][]{botteon24a754, botteon25, balboni25arx, vanweeren25sub}. As a first step, residual radio frequency interference not flagged by the SDP pipeline was removed using \aoflagger\ \citep{offringa10, offringa12}. Low S/N channels at the edges of the band were also discarded.  The data were then averaged using \dpppE\ \citep[\dppp;][]{vandiepen18} and compressed with \dysco\ \citep{offringa16}. Due to the strong ionospheric errors, we performed a series of ``scalarphase+scalarcomplexgain'' direction-dependent (DD) calibration cycles on a large field-of-view (FoV) of $4.17\deg \times 4.17\deg$ using eight directions, partially correcting for the ionospheric distortions. To further improve the calibration, sources outside a reduced FoV of $1.15\deg \times 1.15\deg$ were removed, and a new DD calibration was performed using seven updated directions, successfully correcting for most of the remaining DD effects in the field. Given the relatively small angular extent of our target, we further limited the visibility data to a FoV of $0.26\deg \times 0.26\deg$ to enable faster re-imaging. On this smaller FoV, a final cycle of direction-independent (DI) calibration was carried out, effectively reducing residual artifacts around NVSS J113521$-$201957, a bright radio galaxy belonging to \cl. \\
\indent
Final imaging was done with \wsclean\ v3.6 \citep{offringa14}, adopting a robust weighting of the visibilities of $-0.5$ \citep{briggs95} and the multiscale multifrequency deconvolution algorithm \citep{offringa17}. Our full-resolution image ($9.0\arcsec \times 8.0\arcsec$) has a noise of 4.7 \mujyb. To enhance the S/N of the faint diffuse emission, we also produced source-subtracted images at lower resolution (see Section~\ref{sec:results}). The central frequency of our \meerkat\ images is 822~MHz and the assumed uncertainty on the flux density scale is 15\% \citep[\eg][]{sikhosana25}.

\subsection{uGMRT}

We observed \cl\ with the uGMRT in band 3 (300--500 MHz) for 8 h (project code: 47\_058, PI: T. Caglar) on 2025 January 6. We employed the GMRT Wideband Backend \citep[GWB;][]{reddy17}, spanning 200~MHz of bandwidth, which was covered with 4096 channels. The flux density calibrators 3C147 and 3C286 were observed for 12 min each at the start and at the end of the observation, respectively. \\
\indent
The processing was carried out with the \spamE\ \citep[\spam;][]{intema09} pipeline, which performs instrumental calibration (bandpass and flux density scale), data averaging, flagging, and direction-dependent calibration. The 200~MHz bandwidth was divided by \spam\ into six frequency slices of 33.3 MHz each, which were processed independently. This is because \spam\ was originally developed to processed GMRT Software Backend \citep[GSB;][]{roy10} data, which had a smaller bandwidth. Once calibrated, the six frequency slices were jointly imaged with \wsclean\ similarly as described above for \meerkat. For reference, our full-resolution image ($8.3\arcsec \times 4.5\arcsec$), obtained with a robust weighting of the visibilities of $-0.5$, has a noise of 28 \mujyb. The central frequency of our \ugmrt\ images is 400~MHz and the assumed uncertainty on the flux density scale is 10\% \citep{chandra04}.

\begin{figure*}[h]
 \centering
 \begin{minipage}[b]{0.8\linewidth}
 \includegraphics[width=.95\hsize,trim={0.2cm 0.2cm 0.6cm 0.3cm},clip]{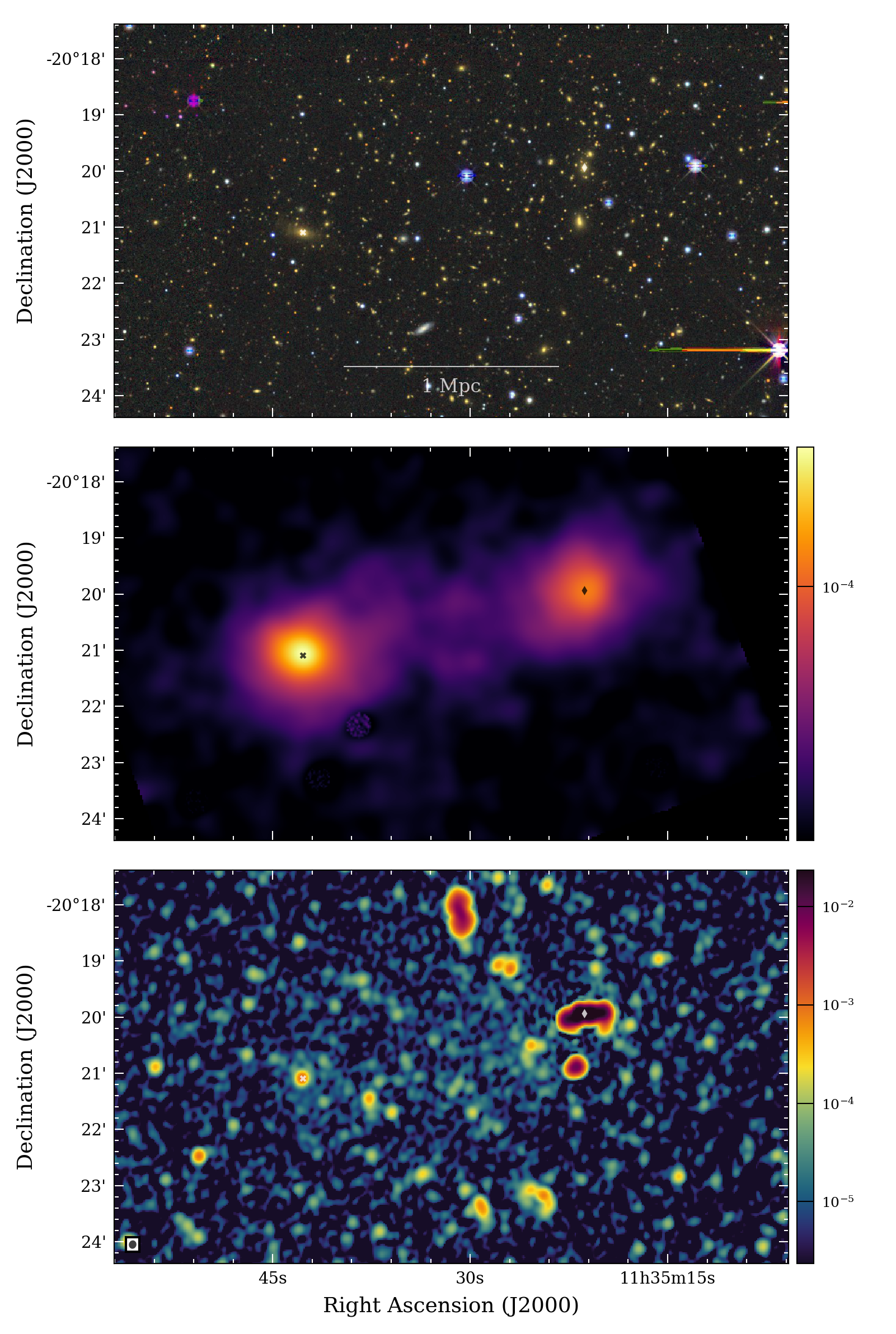}
 \end{minipage}% <- don't forget this %
 \hfill
 \begin{minipage}[b]{0.2\linewidth}
 \caption{Multiband view of \cl. \textit{Top}: optical \textit{g,r,z} image from the DESI Legacy Imaging Surveys \citep{dey19}. \textit{Center}: X-ray adaptively smoothed image in the 0.5--2.0~keV band with point sources removed \citep{doner25sub}. \textit{Bottom}: \meerkat\ radio image at 822~MHz with a beam of $9.0\arcsec \times 8.0\arcsec$ (shown in the bottom left corner) and a noise of $\sigma_{\rm rms} = 4.7$ \mujyb. The cross and diamond markers indicate the position of the galaxies mentioned in the text. The color bar units are counts s$^{-1}$ for the X-ray image and Jansky beam$^{-1}$ for the radio image.}
 \label{fig:multiband}
 \end{minipage}
\end{figure*}

\begin{figure}[h]
 \centering
 \includegraphics[width=.49\hsize,trim={1.8cm 1.1cm 0.2cm 0.5cm},clip]{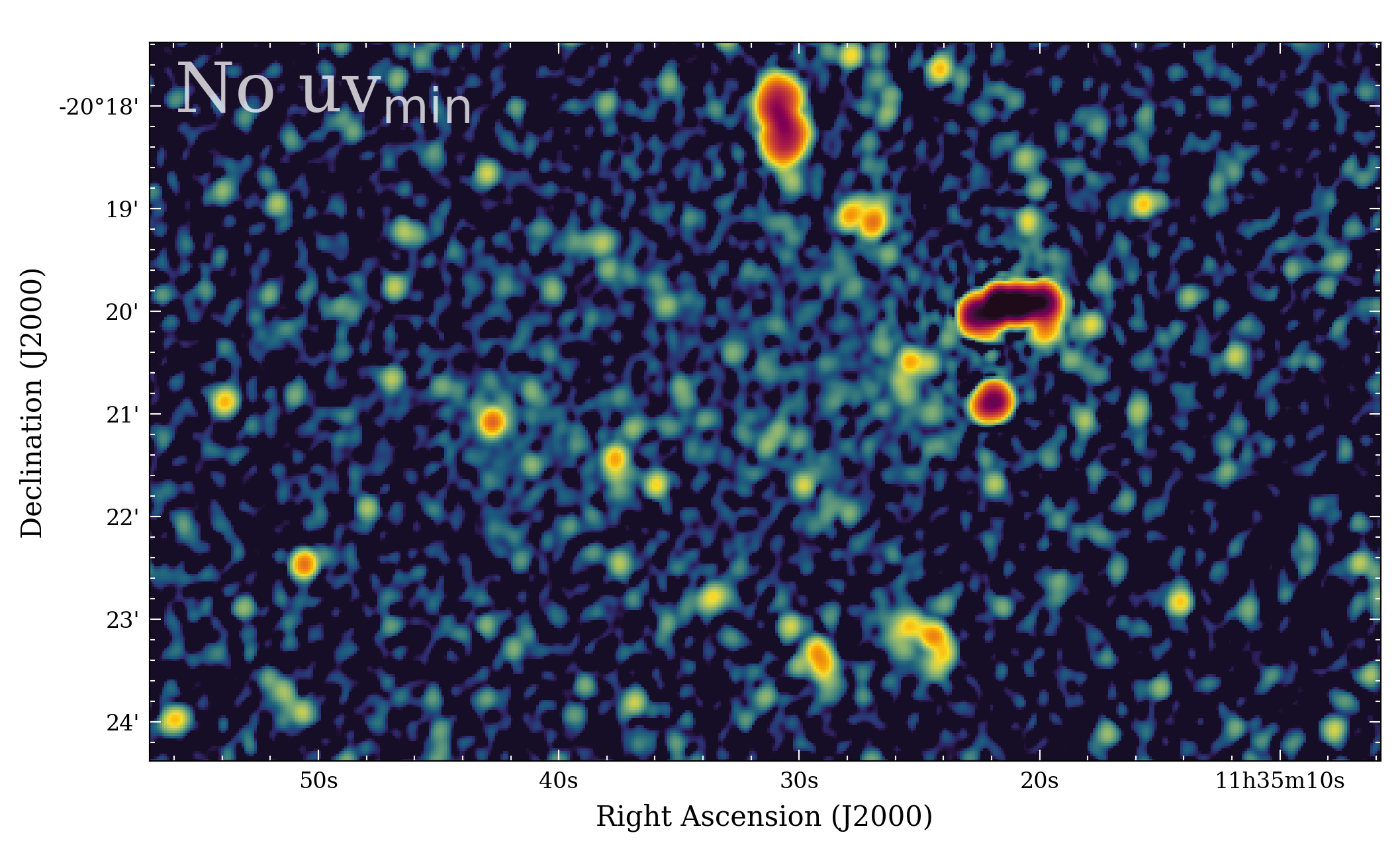}
 \includegraphics[width=.49\hsize,trim={1.8cm 1.1cm 0.2cm 0.5cm},clip]{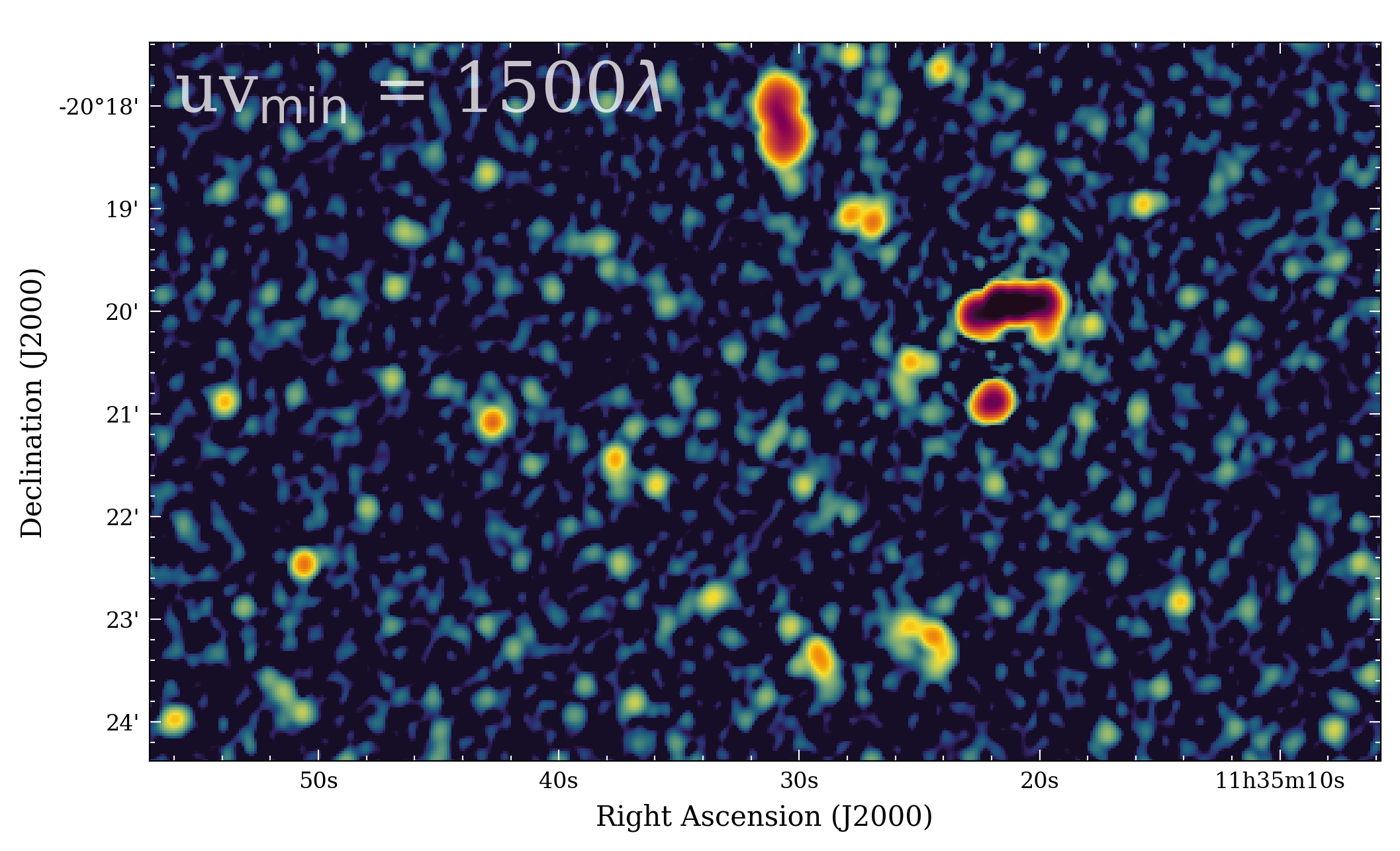}
 \includegraphics[width=\hsize,trim={0.2cm 0.3cm 0.2cm 0.2cm},clip]{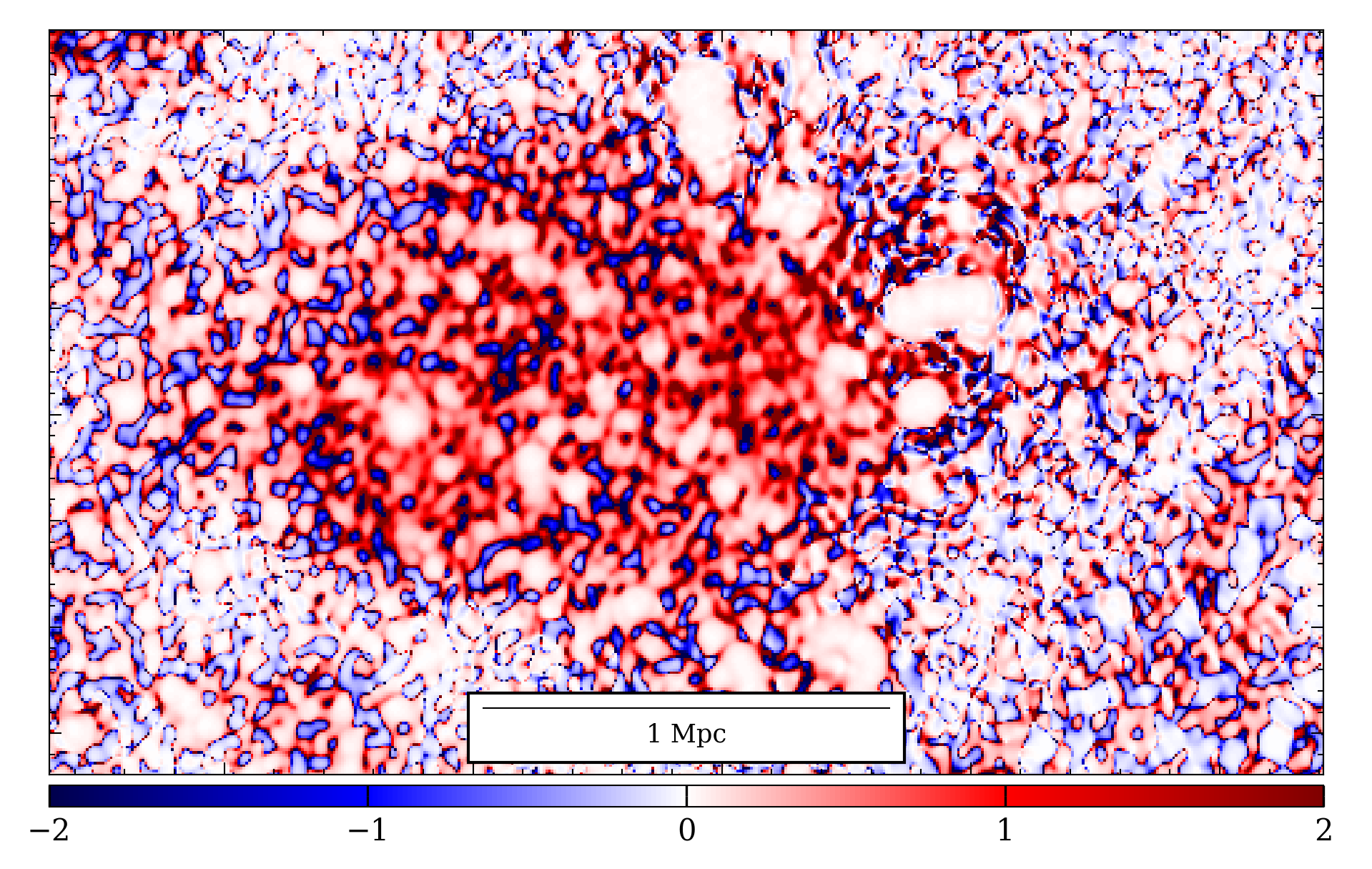}
 \caption{Highlighting the diffuse radio emission excess in \cl. \textit{Top panels}: \meerkat\ images obtained by imaging all baselines (\textit{left}, same as in Fig.~\ref{fig:multiband}) and only baselines longer than $1500\lambda$ (\textit{right}). \textit{Bottom panel}: fractional difference image, computed by subtracting the top right image from the top left image and dividing the result by the former.}
 \label{fig:fracdiff}
\end{figure}

\begin{figure}[h]
 \centering
 \includegraphics[width=\hsize,trim={0.2cm 0.2cm 0.95cm 0.25cm},clip]{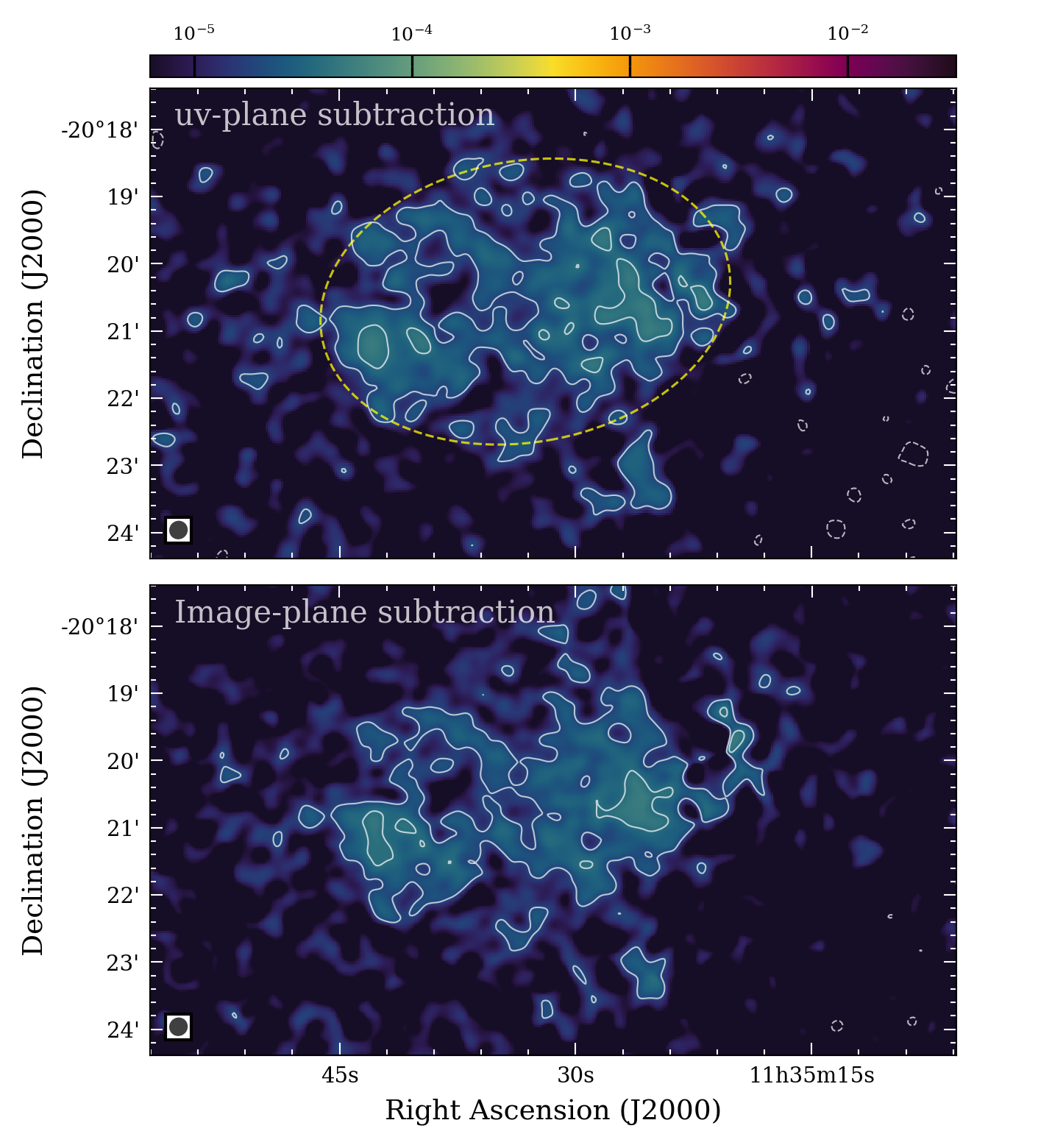}
\caption{Source-subtracted low-resolution \meerkat\ images at 822~MHz. \textit{Top}: image produced by subtracting compact sources in the \uv-plane and reimaging the visibilities with a Gaussian taper, yielding a beam of $16.8\arcsec \times 16.3\arcsec$ and a noise of $\sigma_{\rm rms} = 6.3$ \mujyb. The yellow ellipse denotes the region used to extract the flux densities of the diffuse emission. \textit{Bottom}: image obtained by subtracting compact sources in the image-plane at high resolution, then convolving the result to the same $16.8\arcsec \times 16.3\arcsec$ beam. Solid contours are spaced by a factor of 2 from the $3\sigma_{\rm rms}$ level. The $-3\sigma_{\rm rms}$ contour is shown in dashed. The beams are shown in the bottom left corners. The color bar units are Jansky beam$^{-1}$.}
 \label{fig:lowres}
\end{figure}

\begin{figure}[h]
 \centering
 \includegraphics[width=\hsize,trim={0.2cm 0.2cm 0.25cm 0.5cm},clip]{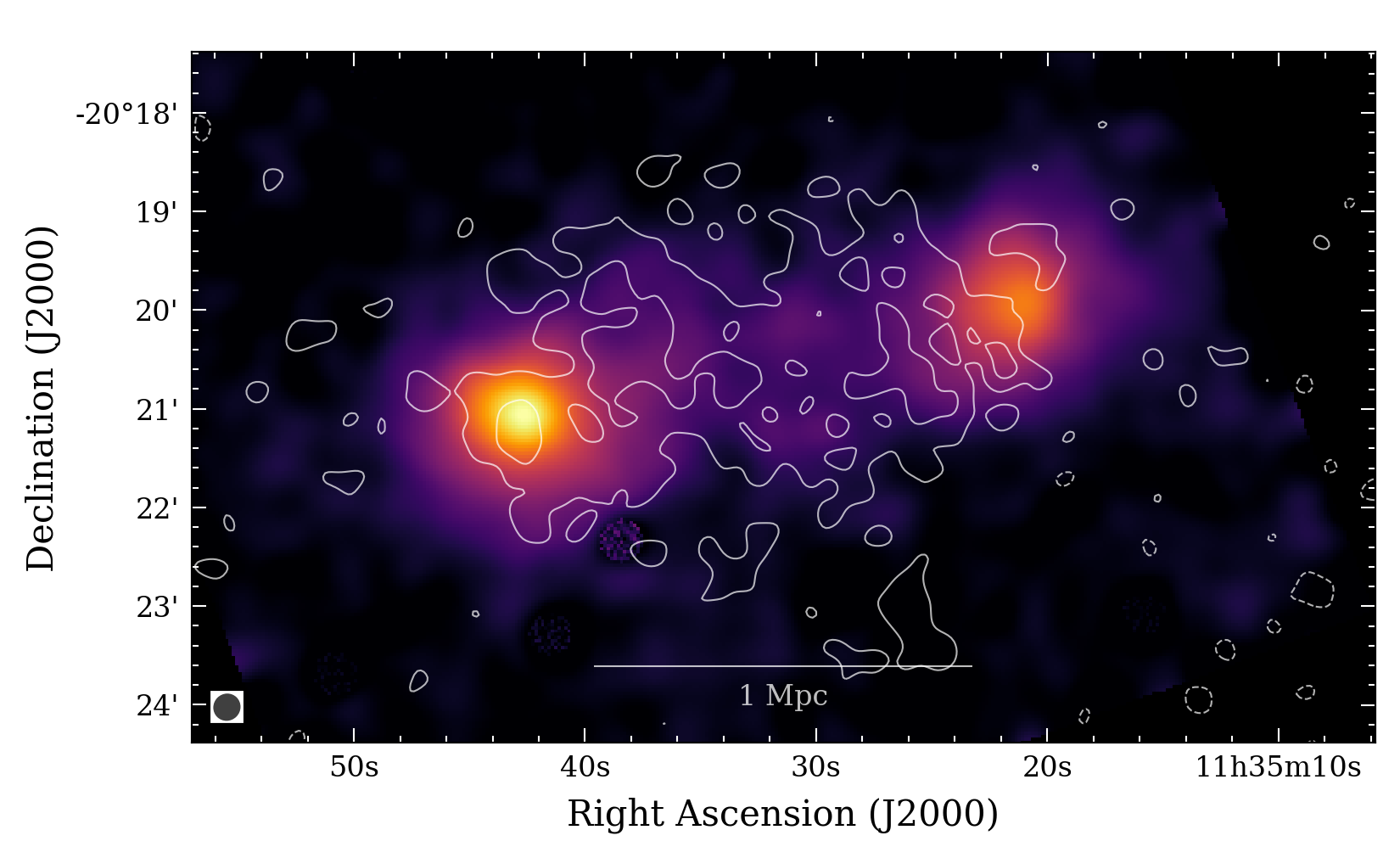}
 \caption{X-ray image with the \meerkat\ low-resolution radio contours with sources subtracted in the \uv-plane overlaid. The beam of the radio image is shown in the bottom left corner.}
 \label{fig:radiox}
\end{figure}

\begin{figure}[h]
 \centering
 \includegraphics[width=\hsize,trim={0cm 0cm 0cm 0cm},clip]{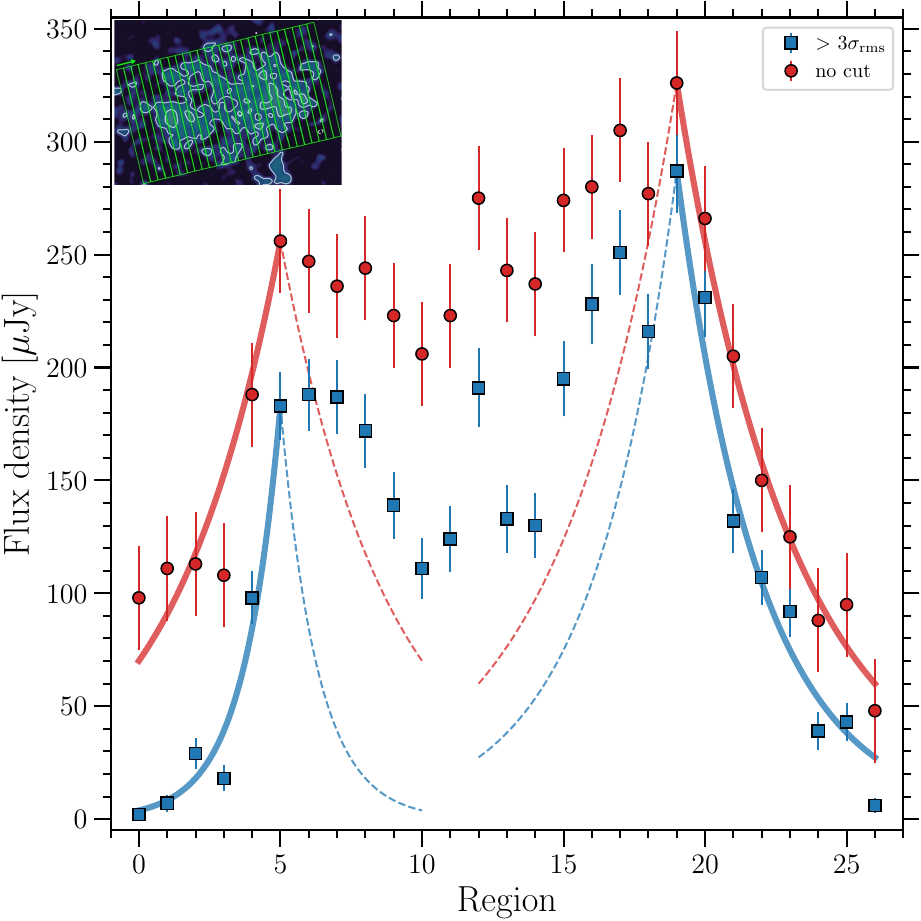}
 \caption{Flux density profile across the cluster pair from east to west integrated above the $3\sigma_{\rm rms}$ level and over the entire rectangular regions (see legend). The solid lines are obtained by fitting an exponential decaying model to the data points on the side away from the opposite cluster, starting from the emission peak. The dashed lines represent the mirrored model toward the other cluster.}
 \label{fig:profile}
\end{figure}

\section{Results}\label{sec:results}

A multiband view of \cl\ is shown in Fig.~\ref{fig:multiband}. In the optical, the eastern subcluster is dominated by the brightest cluster galaxy (BCG), which coincides with the peak of the X-ray emission and hosts a compact radio core (RA: 11$^{\rm h}$35$^{\rm m}$42.69$^{\rm s}$; DEC: $-$20\deg21\arcmin06.21\arcsec, cross marker). The western subcluster appears more optically rich and, at the center of its X-ray emission, hosts the brightest radio galaxy in the field (NVSS J113521$-$201957), which exhibits jets resolved along the east-west direction (RA: 11$^{\rm h}$35$^{\rm m}$21.35$^{\rm s}$; DEC: $-$20\deg19\arcmin56.76\arcsec, diamond marker). Their angular separation of $\sim$308 arcsec corresponds to a projected physical distance of $\sim$1340 kpc at $z=0.29$. \\
\indent
Despite the relatively high angular resolution of the \meerkat\ image in Fig.~\ref{fig:multiband}, there is an indication that low-level diffuse emission is present in the system. We note that this emission was picked up by the model during the deconvolution process, and that the sensitivity of the \ugmrt\ image is significantly lower (Fig.~\ref{fig:gmrt_images}). To assess whether the emission is mainly diffuse or instead due to unresolved point sources, we produced a \meerkat\ image using only baselines longer than 1500$\lambda$. In this image, shown alongside the image without inner \uv-cut in Fig.~\ref{fig:fracdiff} (\textit{top panels}), the extended emission disappears, suggesting that the signal is indeed associated with short baselines and thus originates from diffuse, low surface brightness structure. The bottom panel of the figure shows the fractional residual between the images in the top panels, highlighting the presence of excess diffuse emission in the cluster pair. \\
\indent
Disentangling the contribution of discrete sources to that of the diffuse emission is a crucial but difficult task, and results can significantly change if this step is done in the image-plane or \uv-plane \citep[\eg][]{botteon25, rajpurohit25profiles}. Here, we adopt both approaches to assess the robustness of the procedure. In the top panel of Fig.~\ref{fig:lowres}, we show the result obtained by reimaging the \meerkat\ data at lower resolution after subtracting from the visibility data the model of the discrete sources shown in Fig.~\ref{fig:fracdiff} (top right panel). In the bottom panel of Fig.~\ref{fig:lowres},  we present the result of the image-plane subtraction. This was performed by masking all pixels above $2\sigma$ in the high-resolution image of Fig.~\ref{fig:fracdiff} (top right panel), and subsequently convolving the image to match the resolution of the \uv-plane-subtracted image for a direct comparison. Overall, the morphology of the radio contours is broadly consistent between the two methods, confirming the presence of faint extended emission in the system. Notably, diffuse emission is observed connecting the two clusters, spatially aligned with the X-ray emission bridge (Fig.~\ref{fig:radiox}). This is the only region in the field where there is significant diffuse radio emission in \meerkat, whereas the \ugmrt\ provides a non-detection, as shown in the large-FoV images of Fig.~\ref{fig:large_fov}. The implications of this non-detection on the spectral index of the emission will be discussed at the end of the next Section. \\
\indent
In the following, we adopt the image with source subtraction performed in the \uv-plane, as the use of a Gaussian taper in this approach allows for a more reliable deconvolution of the extended flux. The extended emission projected size is $\sim$1500 kpc $\times$ 800 kpc. Its flux density integrated over the pixels above $3\sigma_{\rm rms}$ ($2\sigma_{\rm rms}$) located within the yellow ellipse shown in Fig.~\ref{fig:lowres} (top panel) is $S = 3.5 \pm 0.1 \pm 0.5$ mJy ($4.4 \pm 0.1 \pm 0.7$ mJy). As an alternative estimate, we extrapolated the average surface brightness of the diffuse emission over the regions where discrete sources were subtracted, obtaining $S = 3.7\pm0.1\pm0.6$ mJy ($4.3\pm0.1\pm0.6$ mJy). The two methods provide consistent results within the statistical errors (reported first). The total error budget is dominated by the systematic uncertainty (reported second). The $k$-corrected radio power (assuming $\alpha=1.3$) at 822~MHz is $P = 1.0 \pm 0.1 \times 10^{24}$ \whz\ ($1.3 \pm 0.2 \times 10^{24}$ \whz).

\section{Discussion}\label{sec:discussion}

The newly discovered extended radio emission connects the two X-ray clumps (Fig.~\ref{fig:radiox}) and exhibits two surface brightness peaks: one in the east, near the peak of the eastern cluster X-ray emission (possibly indicating the presence a radio halo-like source), and another in the west, offset by $\sim$300 kpc from the center of the western cluster X-ray emission. The origin of the latter radio peak is unclear, as it could trace a region where the radio emission is higher either due to the local dynamics of the thermal gas (which may boost the nonthermal emission), or due to the presence of relativistic plasma injected in the medium by cluster radio galaxies. The two peaks can be noticed also in the profiles shown in Fig.~\ref{fig:profile}, which were derived by integrating the flux density within rectangular regions of 16\arcsec\ width along the source extension (see inset). Starting from each peak, we fitted the data points on the side opposite the other cluster with a decaying exponential model and mirrored the resulting profile toward the center of the cluster pair. These fits are not intended to have any physical meaning, but are rather used to visualize that the emission between the clusters declines more gradually, and that it is not simply due to the superposition of the two exponential profiles. This suggests that the radio emission connecting the two clusters has been enhanced during their interaction. \\
\indent
The faint radio emission, which closely follows the X-ray morphology of the thermal gas, likely originates from turbulence in the region between the two interacting clusters. Since it remains uncertain whether \cl\ is observed before or after pericenter passage \citep{doner25sub}, we discuss the interpretation for its origin in both dynamical scenarios. \\
\indent
In the pre-pericenter case, \cl\ would resemble systems such as A399-A401 and A1758N-S, which have been classified as systems that have not yet reached core passage \citep[\eg][]{sakelliou04, david04, hincks22, machado24}. In these cluster pairs, the presence of radio bridges connecting the two subclusters is particularly intriguing, as it implies the activation of mechanisms capable of reaccelerating relativistic particles and amplifying magnetic fields prior to the cluster-cluster collision \citep{brunetti20}. A notable difference between \cl\ and these two cluster pairs is the separation of the subclusters: $\sim$1.3 Mpc for \cl, $\sim$2 Mpc for A1758N-S, and $\sim$3 Mpc for A399-A401. Since these are sky-projected distances, they represent lower limits on the true 3D separations. Among these systems, only for A399-A401 have there been attempts to estimate the actual spatial separation between the two subclusters \citep{akamatsu17filament, bonjean18, hincks22}. In the case of \cl, which is closer to pericenter passage, the true physical separation between the two subclusters is not expected to be significantly larger than the projected value. Estimating the 3D separation between the eastern and western components will require dedicated follow-up observations, particularly spectroscopic optical data to assess the presence of a potential merger component along the line-of-sight. \\
\indent
In the post-pericenter case, the formation of extended radio emission can be more easily explained within the framework of turbulent reacceleration. Indeed, in this case, turbulence is naturally generated in the wake of the interacting clusters \citep[\eg][]{roettiger97, roettiger99magnetic, iapichino08wind}, where it can dissipate part of its energy into nonthermal components, producing diffuse synchrotron emission. Similar phenomena have been observed in radio halos that exhibit surface brightness extensions due to the interaction with subclusters identified either in the optical or X-ray band \citep[\eg][]{kim89, botteon19lyra, pignataro24a2061}. When the X-ray clumps of the interacting clusters are clearly separated and the radio emission lies in the region between them, the diffuse emission has been sometimes referred to as ``radio bridge''. The prime example is represented by the extended emission located between the Coma cluster and the NGC4839 galaxy group \citep[for recent analyses, see][]{bonafede21, churazov21, mirakhor23coma}. As this is also the case in \cl, the radio emission in this system could still be referred to as a radio bridge, even in this dynamical scenario. We therefore suggest that, whenever the term ``radio bridge'' is used, the dynamical stage of the cluster-cluster encounter should be specified, if possible. \\
\indent
Given the small projected separation between the subclusters in \cl\ one might wonder whether the diffuse emission in this system can be decomposed into distinct components (\eg\ halos and bridge), or whether it should be treated as a single, continuous structure. Or, to rephrase it as an interrogative: how far apart must two interacting clusters be, before or after pericenter passage, to be considered independent systems? From an observational point of view, distinguishing the individual components is only possible if the data have sufficient resolution and sensitivity to resolve separate substructures (both morphologically and, eventually, spectrally). We believe that this is not the case for our \meerkat\ data, and therefore would tend to consider the radio emission in \cl\ as a whole. \\
\indent
The subsequent interpretation of the physical properties of the diffuse emission in \cl\ is limited by the unknown 3D geometry of the system and the lack of spectral index information (see the end of this Section for further discussion on the latter). Keeping this limitation in mind, we can adopt the standard assumptions of the source lying in the plane of the sky and a spectral index of $\alpha = 1.3$ to estimate basic physical properties of the source. These can then be compared with those of radio halos and bridges to evaluate potential similarities and differences with these two classes of sources. \\
\indent
For example, assuming an oblate (prolate) ellipsoidal geometry with semi-axes $r_{\rm maj} = 750$ kpc and $r_{\rm min} = 400$ kpc, the volume of emission would be $V = 1.48\times10^{73}$ cm$^3$ ($2.77\times10^{73}$ cm$^3$). Assuming that the radio emission is uniform in the volume, its mean radio emissivity at 822~MHz is $\langle \epsilon \rangle = P / V = 8.58 \times 10^{-43}$ erg s$^{-1}$ Hz$^{-1}$ cm$^{-3}$ ($4.48 \times 10^{-43}$ erg s$^{-1}$ Hz$^{-1}$ cm$^{-3}$). Here we adopted the flux density integrated above the $2\sigma_{\rm rms}$ level ($S = 4.4$ mJy); assuming the the $3\sigma_{\rm rms}$ level measurement ($S = 3.5$ mJy), the emissivity would be $\sim$20\% lower. Extrapolating the results at 1.4 GHz, the emissivity would fall in the range $2.3-4.3 \times 10^{-43}$ erg s$^{-1}$ Hz$^{-1}$ cm$^{-3}$, \ie\ on the low-end of the emissivity distribution of radio halos \citep{murgia24coma}. For comparison, the emissivity estimated for the bridges in A399-A401 and A1758N-S are an order of magnitude lower \citep{govoni19, botteon20a1758}. Interpreting the higher emissivity in \cl\ is challenging given its uncertain merger phase. If the system is observed before pericenter passage, the high value may suggest that radio bridges can span a broad range of emissivities (similar to radio halos) and that this could, for example, depend on the separation between the two clusters. In the post-pericenter scenario, numerical simulations predict that the turbulent energy flux is higher \citep[\eg][]{vazza17turbulence}, which may in turn lead to higher synchrotron emissivities. The \xrism\ satellite is currently probing the velocity dispersion and bulk velocity in the ICM with X-ray microcalorimeter spectroscopy \citep{xrism25a2319, xrism25a2029core, xrism25coma, xrism25a2029beyond, xrism25centaurus}. Although observationally very demanding, extending such measurements to intracluster bridges would be crucial to constrain the level of turbulence and, in turn, understand the efficiency of particle acceleration in these low-density environments. Similarly, following the evolution of relativistic electrons in numerical simulations through tracer particles in the region between colliding clusters, both before and after core passage, would help determining whether the properties of the radio emission in bridges differ between the two dynamical phases. We again remind the reader that the estimates reported above rely on the following assumptions: that the source lies in the plane of the sky, it uniformly fills an ellipsoidal volume, and it has a spectral index of $\alpha = 1.3$. \\
\indent
The power of radio halos is known to strongly correlate with cluster mass. This $P$--$M$ relation has been the focus of several statistical studies, mostly relying on SZ-derived cluster masses from \planck\ \citep[\eg][]{cassano13, duchesne21eor, cuciti21b, cuciti23, vanweeren21}. As \cl\ is a \planck\ cluster, we can assess where it lies in the $P$--$M$ plane by using the most recent relation derived at a frequency close to our \meerkat\ 822~MHz observation; that is, the relation at 1.4~GHz obtained by \citet{cuciti21b}. The extrapolated power of \cl\ at 1.4~GHz is $6.3\times 10^{23}$ \whz, placing it a factor of $\lesssim$2 below the best-fit relation\footnote{We refer to the ``radio halos only'' BCES Y|X fit of the statistical sample.}, in the region mostly occupied by upper limits and ultra-steep spectrum radio halos. It is worth noting that the mass of \cl\ is close to the mass cut of $\mfive > 6 \times 10^{14}$ \msun\ of the sample considered by \citet{cuciti21b}, therefore it lies near the boundary where the relation becomes less constrained. Moreover, the mass estimate from \planck\ for cluster pairs at moderate redshift is limited by the low angular resolution of the instrument, which prevents it from resolving the individual SZ components of closely spaced systems, potentially biasing the mass measurement \citep[see][for a similar argument on A1758N-S]{botteon18a1758}. In these cases, the \planck\ mass is likely more representative of the most massive component, that in \cl\ is the eastern subcluster. Overall, it seems reasonable to conclude that \cl\ falls below the $P$--$M$ relation of radio halos. We note that, in the turbulent reacceleration scenario, underluminous radio halos are generally expected in the early or late stages of cluster mergers \citep{donnert13}. \\
\indent
Lastly, we comment on the non-detection of diffuse emission in the \ugmrt\ data (Figs.~\ref{fig:gmrt_images} and \ref{fig:large_fov}), showing that the emission recovered by \meerkat\ is too faint to obtain a meaningful limit on the spectrum using the available \ugmrt\ observation. As an optimistic estimation, we calculate the upper limit on the flux density at 400~MHz as $S = 2\sigma_{\rm rms} \times \sqrt{N_{\rm beam}}$, where $N_{\rm beam} = A_{\rm source}/A_{\rm beam}$ is the ratio between the area covered by the emission to that of the beam.  We adopt the $2\sigma_{\rm rms}$ level as the threshold, as it represents the limit above which the flux density of an extended radio source can be considered reliable \citep[\eg][]{botteon20a2255}. We assume that the emission covers the same area covered in the \meerkat\ image, adopting an ellipse with $A_{\rm source} = \pi (750 \times 400)$ kpc$^2$. By considering the resolution and noise of the \ugmrt\ image of Fig.~\ref{fig:large_fov} (left panel), the $2\sigma_{\rm rms}$ upper limit on the 400~MHz flux density becomes $S < 0.88 \times \sqrt{248} \approx 13.8$ mJy. This translates to an upper limit on the spectral index of $\alpha \lesssim 1.6-1.8$, depending on whether the integrated flux density at the $2\sigma_{\rm rms}$ or $3\sigma_{\rm rms}$ level in \meerkat\ is used. A more stringent limit at the $3\sigma_{\rm rms}$ level on the 400~MHz flux density would instead imply $\alpha \lesssim 2.2-2.5$. As anticipated, these values, which were obtained using an optimistic approach, are not useful to derive meaningful insights on the spectral index (and thus on the origin) of the diffuse emission. A more rigorous method would require to inject fake radio emission in the visibility data to estimate the upper limit, taking fully into account the flux density losses due to the missing short spacings and sparse \uv-coverage of the interferometric data \citep[see \eg][]{johnstonhollitt17arx, george21ul, bruno23psz2dr2}. We do not  pursue this approach here as it would lead to even looser constraints. The reader is informed that the reported upper limits, despite already being relatively weak, were deliberately derived under favorable assumptions.

\section{Conclusions}\label{sec:conclusions}

We reported on \meerkat\ 822~MHz and \ugmrt\ 400~MHz observations of PSZ2 G279.79+39.09 (\cl), a poorly studied galaxy cluster pair at $z = 0.29$, likely observed close (either before or after) pericenter passage. In X-rays, the two clusters, which are separated by a projected distance of $\sim$1.3 Mpc, are connected by a bridge of thermal gas. Our \meerkat\ radio images revealed the presence of diffuse emission in the pair, while the \ugmrt\ ones resulted in a non-detection due to the lower sensitivity. The emission is stretched along the east-west direction, with a projected size of $\sim$1500 kpc $\times$ 800 kpc and a radio power at 822~MHz integrated above 3$\sigma_{\rm rms}$ (2$\sigma_{\rm rms}$) of $1.0 \pm 0.1 \times 10^{24}$ \whz\ ($1.3 \pm 0.2 \times 10^{24}$ \whz). \\
\indent
From the flux density profile extracted along the cluster pair, we found indication that the emission in-between the two subclusters has been enhanced during the encounter. We discussed the properties of the diffuse radio emission, considered as a whole, in relation to those of radio halos and the radio bridges in A399-A401 and A1758N-S, while acknowledging the uncertainties related to the unknown 3D separation, geometry and spectral index of the emission. In principle, the estimated emissivity and radio power would be in line with that expected by a diffuse emission generated by turbulence in an early phase before or after the collision. However, further observations are needed to confirm the assumptions of our analysis. On the one hand, new radio data are crucial to obtain a reliable measurement of the spectral index, as the upper limits derived from the available \ugmrt\ data are not sufficiently constraining. On the other hand, optical and high-resolution SZ observations would help to clarify the system geometry, enabling a more thorough investigation of the physical properties of the emission. \\
\indent
To conclude, the newly discovered radio emission in \cl\ highlights the presence of nonthermal phenomena in-between a cluster pair, likely driven by turbulence injected during the cluster-cluster encounter. This work contributes to the growing effort to investigate the properties of diffuse radio emission in the Universe beyond the denser ICM regions typically probed by observations by targeting systems with multiple cluster components. This field is likely to flourish with the advent of next-generation facilities, with the \ska\ leading the way.

\begin{acknowledgements}
We thank the referee for the thoughtful comments on the merger scenario and Gianfranco Brunetti, Kosuke Nishiwaki, and John ZuHone for valuable discussion. The \meerkat\ telescope is operated by the South African Radio Astronomy Observatory, which is a facility of the National Research Foundation, an agency of the Department of Science and Innovation.
We thank the staff of the GMRT for support. GMRT is run by the National Centre for Radio Astrophysics of the Tata Institute of Fundamental Research.
This research made use of the LOFAR-IT computing infrastructure supported and operated by INAF, including the resources within the PLEIADI special ``LOFAR'' project by USC-C of INAF, and by the Physics Dept. of Turin University (under the agreement with Consorzio Interuniversitario per la Fisica Spaziale) at the C3S Supercomputing Centre, Italy.
This work made use of the following \textsc{python} packages: \texttt{APLpy} \citep{robitaille12}, \texttt{astropy} \citep{astropy22}, \texttt{CMasher} \citep{vandervelden20}, \texttt{matplotlib} \citep{hunter07}, and \texttt{numpy} \citep{vanderwalt11}.
%\texttt{scipy} \citep{virtanen20} \texttt{iminuit} \citep{dembinski24}
\end{acknowledgements}


\bibliographystyle{aa}
\bibliography{library.bib}

\begin{appendix}

\onecolumn

\section{Additional radio images}\label{app:additional_images}

\begin{figure}[h]
 \centering
 \includegraphics[width=.5\hsize,trim={0.2cm 0.2cm 0.4cm 0.3cm},clip]{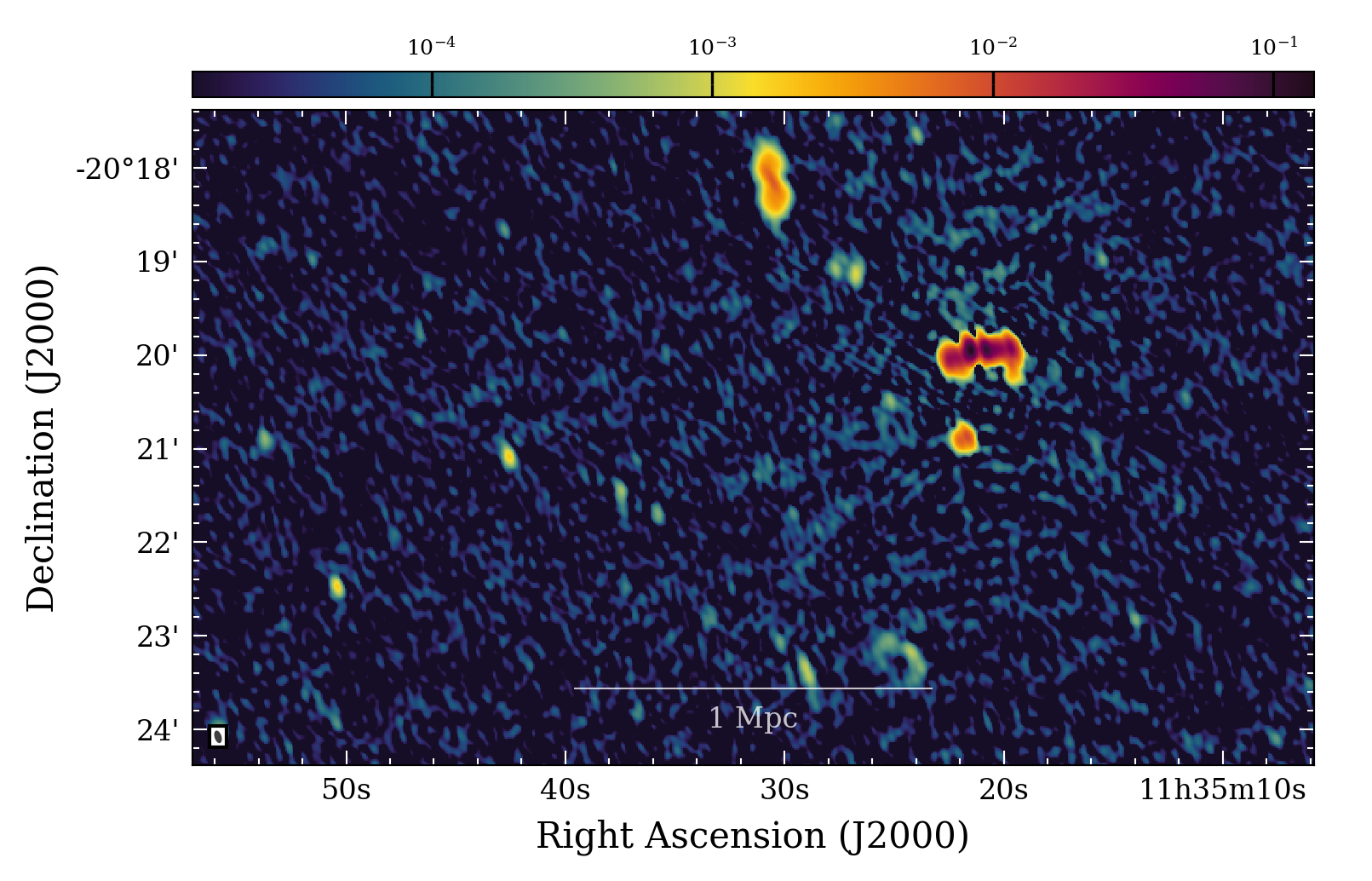}
 \caption{\ugmrt\ radio image at 400~MHz with a beam of $8.3\arcsec \times 4.5\arcsec$ (shown in the bottom left corner) and a noise of $\sigma_{\rm rms} = 28$ \mujyb. The color bar units are Jansky beam$^{-1}$.}
 \label{fig:gmrt_images}
\end{figure}

\begin{figure*}[h]
 \centering
 \includegraphics[width=\hsize,trim={0cm 0cm 0cm 0cm},clip]{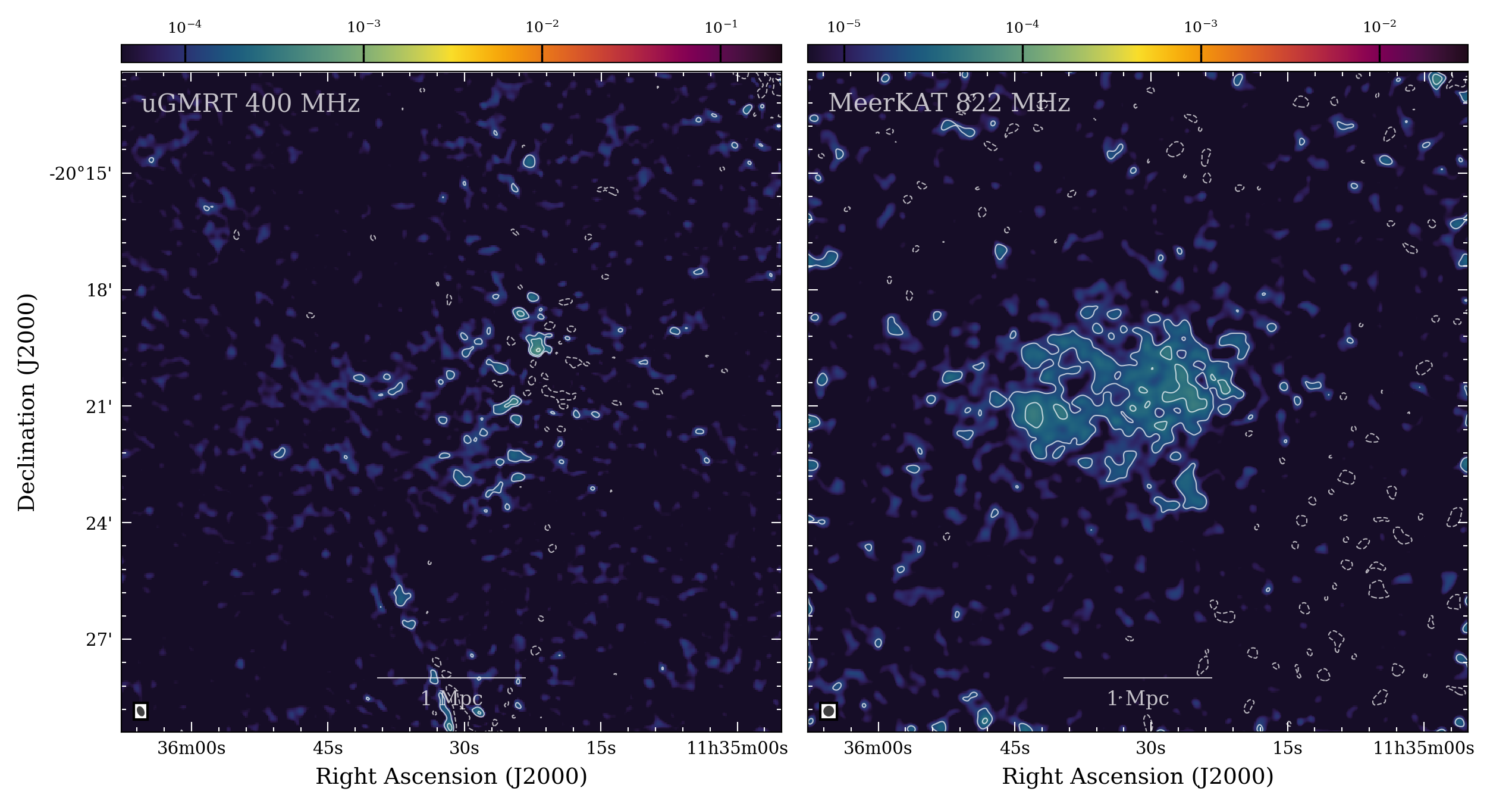}
 \caption{Large-FoV source-subtracted images of \cl. \textit{Left}: \ugmrt\ 400~MHz image at $16.4\arcsec \times 10.8\arcsec$ resolution with a noise of 44 \mujyb. \textit{Right}: \meerkat\ 822.MHz image at $16.8\arcsec \times 16.3\arcsec$ resolution with a noise of 6.3 \mujyb\ (the same of Fig.~\ref{fig:lowres}, top panel). Solid contours are spaced by a factor of 2 from the $3\sigma_{\rm rms}$ level. The $-3\sigma_{\rm rms}$ contour is shown in dashed. The beams are shown in the bottom left corners. The color bar units are Jansky beam$^{-1}$.}
 \label{fig:large_fov}
\end{figure*}

\end{appendix}

\end{document}